\documentclass[10pt]{article}
\pdfoutput=1
\usepackage{jheppubsnowmass}
\usepackage{amsmath,amssymb,amsfonts,graphicx,slashed,color,amsthm, mathtools, enumerate, tensor}
\usepackage{subcaption}
\usepackage{setspace}
\usepackage[export]{adjustbox}
\usepackage[dvipsnames]{xcolor}
\graphicspath{{pics/}}

\allowdisplaybreaks

\colorlet{darkblue}{blue!70!black}

\colorlet{darkgreen}{green!50!black}
\colorlet{darkred}{red!50!black}

\newcommand{\mc}{\mathcal}

\def\bea{\begin{eqnarray}}
\def\eea{\end{eqnarray}}
\def\be{\begin{equation}}
\def\ee{\end{equation}}

\title{
Snowmass White Paper:\\
Cosmology at the Theory Frontier
}

\author[a]{Raphael Flauger,}
\author[b]{Victor Gorbenko,}
\author[c]{Austin Joyce,}
 \author[d]{Liam McAllister,}
 \author[e]{Gary Shiu,}
 \author[f]{\\Eva Silverstein}
  \affiliation[a]{Department of Physics, UC San Diego, La Jolla, CA, 92093}
 \affiliation[b]{Laboratory for Theoretical Fundamental Physics, Institute of Physics,
\'{E}cole Polytechnique F\'{e}d\'{e}rale de Lausanne, Switzerland}
\affiliation[c]{Kavli Institute for Cosmological Physics, Department of Astronomy and Astrophysics, University of Chicago, Chicago, IL 60637}
 \affiliation[d]{Department of Physics, Cornell University, Ithaca, NY 14853}
 \affiliation[e]{Department of Physics, University of Wisconsin-Madison, Madison, WI 53706}
 \affiliation[f]{Stanford Institute for Theoretical Physics, 382 Via Pueblo, Stanford, CA 94305}

 \vspace{5mm}

\vspace{1cm}

\emailAdd{flauger@physics.ucsd.edu}

\emailAdd{victor.gorbenko@epfl.ch}

\emailAdd{austinjoyce@uchicago.edu}
\emailAdd{mcallister@cornell.edu}
\emailAdd{shiu@physics.wisc.edu}
\emailAdd{evas@stanford.edu}

\abstract{

The precision cosmological model describing the origin and expansion history of the universe, with observed structure seeded at the inflationary cosmic horizon, demands completion in the ultraviolet and in the infrared.
The dynamics of the cosmic
horizon also suggests an associated entropy, again requiring a microphysical theory.  Recent years have seen enormous progress in understanding the structure of de Sitter space and inflation in string theory, and of cosmological observables captured by quantum field theory and solvable deformations thereof.
The resulting models admit ongoing observational tests through measurements of the cosmic microwave background and large-scale structure, as well as through analyses of theoretical consistency by means of thought experiments.  This paper, prepared for the TF01 and TF09 conveners of the Snowmass 2021 process, provides a
synopsis of this important area, focusing on ongoing developments and opportunities.

}

\begin{document}

\maketitle
\parskip=10pt

\section{Introduction} \label{sec:intro}

The observed universe is very accurately described by a concordance cosmology in which inflation prepares the initial conditions for the hot Big Bang and $\Lambda$CDM cosmology governs the subsequent evolution.
The development of this precision model involved
evolution from a well-motivated and transformative theory of the quantum origin of structure  \cite{Guth:1980zm,Starobinsky:1980te, Linde:1981mu,Albrecht:1982wi, Mukhanov:1981xt}, through detailed theoretical predictions for observables  \cite{1987MNRAS.226..655B}, to observations confirming the predictions, most recently by the Planck mission \cite{Planck:2018vyg,Planck:2019nip}.
The concordance cosmology is a remarkably deep and successful phenomenological model, but it is not yet a microphysical theory.
Instead, it is a starting point for more complete studies of early and late universe physics, whose observables are ultraviolet-sensitive, i.e.~dependent on high-energy physics \cite{Kachru:2003sx}, and are also dynamically rich in the infrared, i.e.~at low energies and late times.
One of the cardinal tasks for any complete formulation of quantum gravity is to give a microphysical explanation of the accelerating expansion of the universe.
The potential energy landscape of string theory provides a framework for this problem, and recent years have seen significant progress, though not yet a complete solution.

Many reviews and white papers --- see e.g.~\cite{Baumann:2014nda, Silverstein:2016ggb, Silverstein:2017zfk} and \cite{CMB-S4:2016ple,Shandera:2019ufi,Meerburg:2019qqi,Slosar:2019gvt,Abazajian:2019eic,NASAPICO:2019thw,LiteBIRD:2022cnt} --- recount aspects of research at the interface of modern string theory, quantum field theory, and cosmology, but none is entirely up to date in this rapidly evolving field.  In this white paper we
summarize the status of this important area along several axes, focusing on exciting recent advances and future prospects.

The rich and highly structured landscape of string theory yields well-developed mechanisms for de Sitter spacetime, dark energy, and early universe inflation.  This connection introduces meaningful, but model-dependent, observational tests of ideas from string theory.  Recent advances include new classes of compactifications exploiting the rigidity properties of generic internal manifolds, as well as derivation of explicit control parameters in these and previous de Sitter models.
Below we detail these developments and the new opportunities they raise. We also highlight key features of standard models of de Sitter space, large-field inflation, and small-field inflation in string theory.  In the process, we emphasize the opportunities in axion physics broadly construed, both because of exciting near-future observational opportunities and because of the apparent universality of axions within string theory.

Formal low energy effective field theory (EFT) has also seen advances
in characterizing infrared physics in the accelerating universe, including nonperturbative aspects of the wavefunction and probability distribution for primordial perturbations, such as
the strength of the tails.  At the same time, elegant calculations and constraints on correlators related to amplitude and bootstrap theory help to organize the observables, particularly in highly symmetric versions of inflation.  Moreover, deformations of quantum field theory involving $T\bar T$ \cite{Zamolodchikov:2004ce} and generalizations extend the solvable theory space relevant for holographically formulating cosmological spacetimes.

In addition to making the connection to real observables, these  developments also yield concrete progress on the abstract problem of cosmological quantum gravity.
Advances here include four-dimensional de Sitter uplifts of AdS/CFT, emergent geometry and microstate counts from a solvable generalization of the $T\bar T$ deformation, and new handles on Euclidean quantum gravity effects.

The observational prospects remain bright.  Cosmic microwave background (CMB) polarization studies are progressing in space and on the ground, as are large-scale structure (LSS) surveys, bolstered by the latest astrophysics decadal process.  The latter yields a compelling theoretical challenge to extract primordial information from LSS using effective theories, simulations, and numerical methods such as machine learning (ML).   This program includes following through on standard sources of non-Gaussianity in low point correlators, such as the equilateral shape, as well as nonperturbative aspects related to higher $n$-point functions.

Finally, we note that even in this arena focused on the theoretical underpinnings of subtle cosmological and abstract quantum gravity observables, connections to industry are possible.  A growing set of numerical results appear in the current literature, applying ML to metrics and more general partial differential equations arising in string compactifications, and applying insights from cosmological models into machine learning itself, e.g.~in optimization.

\section{de Sitter space and inflation in string theory}\label{sec:string-inflation}

\subsection{General structure}\label{sec:general}

String/M-theory exhibits a complex configuration space connecting regions with varying dimensionalities $D$, topologies,
geometries, and quantized fluxes. To model cosmology, we are interested in dimensional reduction to four dimensions, with weak curvature yielding approximate Einstein gravity and dynamics consistent with the observed $\Lambda$CDM cosmology. The rich but constrained structure of the string/M-theory configuration space yields a number of consistent mechanisms for de Sitter and early universe inflation, some with novel signatures testable with CMB and LSS data.
Moreover, there exist solutions providing direct uplifts of AdS/CFT systems, giving insight into the microphysics of de Sitter quantum gravity.

String theory famously contains gravity.
Essentially as universal is the presence of axions \cite{Peccei:1977hh, Peccei:1977ur, Wilczek:1977pj} in the
four-dimensional effective theories arising from string theory.
Axion fields descend either directly from the $D>4$ dimensional theory, or via nontrivial internal topology threaded by
gauge fields of appropriate rank.  Axions feature in all known weakly-interacting limits of the theory  --- regardless of model-dependent properties such as the presence or absence of supersymmetry at low energies
--- and admit observational testing in diverse parameter regions and datasets. Axions have an underlying period that is small in Planck units \cite{Banks:2003sx},
but couple to fluxes and branes so as to exhibit a monodromy-extended potential with a branched structure.  Compactifications with rich internal topology give rise to a large number of axions (cf.~\cite{Arvanitaki:2009fg}), with a multi-axion potential determined by such couplings.  In the early universe these fields provide natural inflaton candidates, and in the late universe they provide dark matter candidates, as well as the possibility of incorporating the QCD axion to solve the strong CP problem.  In many cases they may even be ultralight in the late universe, yielding additional phenomenological effects.

The effective potential for scalar and pseudoscalar
{(i.e.,~axion)} fields in four dimensions is a central object for studying cosmology in string theory. For a $D$-dimensional warped product
\begin{equation}\label{eq:warped-met}
    ds^2 = u(y) ds^2_{4} + g^{(D-4)}_{ij}(y) dy^i dy^j\,,
\end{equation}
the effective potential in the four-dimensional Einstein frame takes the form \cite{Douglas:2009zn}
\be\label{VeffEframe-string}
V_{\text{eff}}\left(g^{(D-4)}, \dots\right) =\frac{\ell_{D}^{D-2}}{2 G_N^2 \langle u_c \rangle^2} \int\,d^{D-4}y\, \sqrt{g^{(D-4)}}e^{-2\Phi}u_{c}^2\left(-R^{(D-4)}-\frac{1}{4} \ell_{D}^{D-2} {T}^\mu_\mu
-3\left(\nabla u_c/u_c\right)^2\right)
\,,
\ee
in terms of the solution $u=u_c$ of the internal constraint equation
\begin{equation}\label{Constraint}
    \nabla^2 u -\frac{1}{3}\left(R^{(D-4)}+\frac{1}{4}\ell_{D}^{D-2} {T}^\mu_\mu
    \right)u=\frac{C}{6}\,,
\end{equation}
where $C$ is a constant, $\ell_D$ is the $D$-dimensional Planck length, $T_{\mu\nu}$ encodes the stress-energy sources in the theory, and we have defined
\be
\langle u_c \rangle := \int d^{D-4} y \sqrt{  g^{(D-4)}} e^{-2\Phi} u_c\,.
\ee
A maximally symmetric solution to all of the equations of motion yields $V_{\text{eff}}=C/4G_N$.
The expression \eqref{VeffEframe-string} is applicable in
perturbative string theory regimes, for which
we have pulled out the dependence on the string coupling $g_s=e^\Phi$. In the M-theory limit of 11-dimensional supergravity, the formula is the same but without the $e^\Phi$ factors.

The sources of stress-energy appearing in \eqref{VeffEframe-string}-\eqref{Constraint}, as well as the
the internal curvature and warp factor gradients, dilute at large radius and weak coupling \cite{Dine:1985kv}, requiring at minimum a three-term structure in those runaway directions in field space \cite{Silverstein:2001xn}.  Most contributions to the potential are positive, with the leading (tree-level and least dilute) source in $T^\mu_\mu$ being the term proportional to $D-D_{c}$ \cite{Myers:1987fv, deAlwis:1988pr, Silverstein:2001xn, Maloney:2002rr, Friess:2005be}, with $D_c$ the critical dimension.
The next-to-leading terms in the potential are the (generically negative) contributions of the internal curvature and warp factor variations, followed by the effects of branes and fluxes.  Negative intermediate terms in the expansion about large radius and weak coupling play a key role, and emerge from orientifold planes or quantum effects (e.g.~perturbative Casimir energy \cite{DeLuca:2021pej} or nonperturbative instanton effects \cite{Kachru:2003aw}).
The solutions of interest arise when terms at different orders in the expansion can compete, which requires a large or small number in the system, for example a large ratio of integers characterizing the topology or the configuration of quantized fluxes (see e.g.~\cite{Demirtas:2021nlu}).
In cases where quantum effects govern the vacuum structure, the stability of the extra dimensions of string theory proves analogous to the stability of ordinary matter and of degenerate stars, both of which depend on quantum mechanics, as does the simplest theory of the origin of structure in the universe \cite{Mukhanov:1981xt}.

Scalar and pseudoscalar fields in the four-dimensional effective theory descend from geometric deformations $\delta g_{ij}$, while
pseudoscalars descend from potential fields $C^{(p)}$ \cite{Polchinski:1998rr}; collective coordinates of branes, if present, also yield scalars or pseudoscalars.
Geometric deformations enjoy several stabilizing features:  the warp factor buttresses the conformal mode of the internal metric, as explained in \cite{Douglas:2009zn}, and
many compactification manifolds are negatively curved and rigid, exhibiting a strong positive Hessian for metric deformations \cite{Besse:1987pua,DeLuca:2021pej}, along with simple mechanisms for stabilizing the overall volume.
Other well-studied cases such as Calabi-Yau compactifications exhibit \emph{moduli}, i.e.~scalar fields that are massless at tree level.  Such compactifications are amenable to treatment by means of algebraic geometry, in part as a consequence of the supersymmetry that they preserve.  Moduli in Calabi-Yau compactifications can be stabilized via a combination of fluxes \cite{Giddings:2001yu}\
and branes and nonperturbative effects \cite{Kachru:2003aw,Balasubramanian:2005zx}.

The pseudoscalar contribution to the potential arises via the structure of fluxes in string theory, and the corresponding brane sources.  Fluxes $F_{r+1}= dC_{r}$ appear in combinations of the form $\tilde F_{p+1} = F_{p+1} - C_q\wedge F_{p+1-q}$.  The $T^\mu_\mu$ term in
\eqref{VeffEframe-string} includes a contribution from the squared generalized flux $\tilde F\wedge \star \tilde F$, leading to a direct dependence on the axions $c=\int_{\Sigma} C$ threading cycles $\Sigma$.  Importantly, higher powers of generalized fluxes are suppressed at weak coupling and large radius,\footnote{In addition to the dilution in space, Ramond-Ramond fluxes appear with a power of the string coupling.}
and hence are negligible over a large, super-Planckian field range in many models.  Although the leading axion dependence in the potential
appears quadratic, it very often flattens out due to the (frequently calculable) backreaction of the axion energy on metric components \cite{Dong:2010in,McAllister:2014mpa}. This backreaction affects the masses of other degrees of freedom, but basic calculations in the controlled regime of large radius in various models show that this can be a small effect over the $\mathcal{O}(10)$ Planck units required to model early universe inflation.\footnote{We review particular examples of this shortly.}
Similar comments apply to the brane couplings of axions. Nonperturbative effects, on the other hand, contribute oscillatory axion-dependent terms in $V_{\text{eff}}$; these are small corrections in many parameter regimes.

The methods used to achieve theoretical control in this class of problems are standard in theoretical physics.  Tools include perturbation theory (including large-$N$ and large-$D$ expansions), instanton calculus, mathematical theorems (relevant examples include the existence of the Calabi-Yau metric and the rigidity of the explicitly known hyperbolic metric), worldsheet methods (an example being the timelike linear dilaton solution, exact on the worldsheet and tachyon-free and ghost-free for the worldsheet-supersymmetric theory in $10+16k$ dimensions), and numerical methods.
Additional tools include supersymmetric constraints and integrability theory (as in the $T\bar T$ deformation) where applicable.  These latter tools sometimes enable theoretical control of deformations extending far in coupling space, keeping track of the effect on energy levels and state counts, which is particularly useful for conceptual problems and dualities, as we discuss in the cosmological context in \S\ref{sec:holography}.   However, modeling de Sitter space and inflation in the effective, weakly-coupled description involves no such extrapolation, and requires only perturbative control.

\subsection{Power-law stabilization}\label{sec:power-law}

In the regime of string/M theory near weak coupling and large radius, the contributions to the potential energy in
\eqref{VeffEframe-string} organize themselves
into expansions in powers of coupling(s) and inverse radii.  From the general setup of \S\ref{sec:general}, one finds subsets of terms that consistently support accelerated expansion. One example appeared soon after the discovery of the cosmological constant and the observation that string theory contains a `discretuum' of potential energy functions \cite{Bousso:2000xa}.  The leading positive potential term $\propto D-D_c$, an exponentially strong negative contribution from orientifolds, and flux suffice to stabilize an asymmetric orbifold of supercritical string theory in a way amenable to a large-$D$ expansion in which the negative contribution scales exponentially in $D$ \cite{Silverstein:2001xn,Maloney:2002rr}.  This vacuum depends on a balance of forces including a
flux-squared
term competing with the leading string-scale scalar potential, which is justified provided higher orders in flux are either suppressed at large $D$ (as suggested by preliminary calculations \cite{Huajia:largeD}) or in any case contribute net positively.   Intervening years have witnessed tightening constraints on low-energy supersymmetry and on deviations from a positive cosmological constant.
Advances on the theoretical side include the emergence of
explicit interpolations between string backgrounds of different dimensionality, connecting $D>D_c$ and $D=D_c$ (e.g. \cite{Hellerman:2006ff,McGreevy:2006hk}), as well as
further development of large $D$ methods in general relativity and string theory.  The supercritical mechanism remains viable both observationally and theoretically, warranting further investigation of the behavior of the theory at large $D$, which is an evidently generic parametric limit.

Several classes of models appear at $D=D_c$, and the subject has been developed in a number of works as referenced for example in \cite{Silverstein:2016ggb, Silverstein:2017zfk}.\footnote{Errors in some significant details of \cite{Silverstein:2007ac} were discovered by  \cite{Gur-Ari:2013sba}, though the general framework and methods remain valid.}   An approach that incorporates strong
internal gradients is \cite{Cordova:2019cvf} and predecessors.  The essence of the idea is to consider a five-dimensional system that would have a single unfixed scalar $\sigma$ subject to a runaway potential $V(\sigma)\sim \exp(\sigma/M_{\text{pl}})$.  The internal
profile
satisfies an ODE with solutions $\sigma(x^5)$
that approach singularities at two ends of an interval. If orientifold planes can consistently intervene before the singularity, one obtains a de Sitter solution.

The second-strongest positive term in the potential is the curvature term.  Almost every manifold is negatively curved, as is evident already in dimension 2, and this feature appears in the models just discussed \cite{Cordova:2019cvf}\ as well as in \cite{Saltman:2004jh} and in predecessors in the EFT literature \cite{Kaloper:2000jb}.  Above dimension 2, such manifolds are rigid \cite{Besse:1987pua}, as mentioned above. This fact underlies the newest class of de Sitter models \cite{DeLuca:2021pej} in M-theory on irreducible  hyperbolic 7-manifolds, dressed with varying warp and conformal factors and with cusps Dehn-filled as in \cite{Anderson:2003un}, and including 7-form magnetic flux.
In these models the automatically-generated Casimir energy contributes the intermediate negative term in the overall volume stabilization.  The net curvature term in
the M-theory analogue of
\eqref{VeffEframe-string}
scales as the inverse square of the curvature radius, and admits tuning to a small value via explicit discrete moves in concrete hyperbolic manifolds (see \S4.1 of \cite{DeLuca:2021pej}). These features open up the possibility of parametric control, in a system simpler than the closest such analogue, which is string theory on a product of Riemann surfaces \cite{Saltman:2004jh}.  The construction of \cite{DeLuca:2021pej}
can be viewed as a $\mathrm{dS}_4$ uplift of the M2-brane AdS/CFT duality pair, with the internal $S^7$ replaced by $\mathbb{H}_7/\Gamma$, providing a simpler and more realistic example in the spirit of the explicit $\mathrm{dS}_3$ models \cite{Dong:2010pm}\ that uplift the D1-D5 system.

The associated literature exhibits transitions indicating the connectivity of the string/M-theory configuration space.  In string theory such transitions arise via condensation of winding modes \cite{Adams:2005rb}, and it would be very interesting to extend this phenomenon to M-theoretic regimes.  In the same vein, there are concrete calculations relating the two leading terms in the potential via a natural generalization of T-duality \cite{McGreevy:2006hk}. Since $D>D_c$ and negative curvature with its infinite sequences of topologies comprise generic parametric limits of the theory, such connections deserve much further study.

Conversely, it is straightforward to write down string/M-theory backgrounds with limited stress-energy and geometric features that do \emph{not} describe accelerated expansion.  In this vein, partial no-go theorems such as \cite{Maldacena:2000mw,Hertzberg:2007wc} can be useful if not overinterpreted.  However, neither these nor their descendants in the literature incorporate the leading positive terms in the potential, the full complement of known branes and orientifold planes, or Casimir energy.  As such, the configurations governed by such no-go theorems are manifestly non-generic, even within the class of
power-law stabilization constructions.
It is important to recognize that more generic ingredients in fact often simplify aspects of the analysis, enabling potential terms that contribute useful forces,
and introduce rigidity of the internal dimensions. Thus, existing no-go theorems do not have significant implications concerning the existence of power-law stabilized accelerated expansion in the framework of string theory, nor do they currently impact observational predictions.

Next, let us note that the power-law stabilization setting has provided some of the simplest examples of axion monodromy, such as those in \cite{Marchesano:2014mla,McAllister:2014mpa,DeLuca:2021pej}.
These scenarios illustrate the effect of backreaction on metric components, which flattens the potential to sub-quadratic powers. This backreaction also adjusts the mass spectrum of other fields, but these fields do not become light over the inflationary field range, as was specifically checked in the relevant literature starting from the original models.  As a simple example, consider the $V_{\text{inflaton}}\propto \phi^{2/3}$ model summarized in equations (4.10) and (4.13) of   \cite{McAllister:2014mpa}.
In this example, the underlying axion $b$ is tied by the dynamics (backreaction) to the size $L$ of the space as $b\propto 1/L^4$, and is related to the canonically-normalized inflaton field by $\phi\propto b^{3/2}$.  During inflation, $\phi$ evolves from $10 M_{\text{pl}}$ to $M_{\text{pl}}$, rescaling by a factor of $1/10$.  Hence $b$ rescales by a factor of $10^{-2/3}$ and $L$ rescales by a factor of $10^{1/6}$.  The change in $L$ changes the Kaluza-Klein masses, since $M_{\text{KK}}\propto 1/L$, but rescaling these those by a factor of $10^{-1/6}$ is unimportant in the dynamics.
We stress that the leading backreaction of the inflaton on the other scalars, on the (flattened) shape of the potential, and on the spectrum is included in the analysis of the model --- the existence of this backreaction does not invalidate the model.  Quite the opposite, it explicitly illustrates why so-called plateau potentials arise from the effects of the UV completion, as predicted in \cite{Dong:2011uf}.  In the large radius and weak coupling regime of the model, instanton effects are suppressed.
See also \cite{McAllister:2008hb,Flauger:2009ab} and \S\ref{sec:NP} below for a class of examples in nonperturbatively-stabilized spaces where the analogous backreaction effect can be negligible.

Generically there would be multiple fields, leading to a distribution of predictions for the tilt of the power spectrum \cite{Wenren:2014cga}.  Moreover, additional effects studied recently in \cite{DAmico:2021vka,DAmico:2021zdd,DAmico:2022gki}\ enrich the model space in a way that relates to the Hubble tension.  Analyzing this systematically in the top-down multifield context is a very interesting task for the future.

To conclude this discussion of power-law stabilization, let us note additional future directions.  The general expression for the effective potential \eqref{VeffEframe-string} \cite{Douglas:2009zn} in the context of explicit internal manifolds (e.g.~those built from polygons, such as \cite{italiano2020hyperbolic})
enables more general analysis, both analytic and numerical, exploiting the known metric, the rigidity properties, and the parametric limits obtained from sequences of manifolds (e.g.~covers).  One can formulate the slow roll figures of merit by taking functional derivatives of $V_{\text{eff}}$ with respect to the fields and descending in that landscape.  Mathematical theorems concerning systolic geometry and its relation to volume \cite{AgolInbreeding,Thomsonethesis} suggest interesting energy theorems in this context.  These directions interface with the material below in \S\ref{sec:connections}.

\subsection{Nonperturbative stabilization}\label{sec:NP}

Calabi-Yau threefolds are a famous class of vacuum solutions of string theory \cite{Candelas:1985en} that have been a wellspring of ideas in mathematics and theoretical physics.
The richness of Calabi-Yau geometry stems in part from special holonomy and the corresponding supersymmetry.  Moreover, Yau's proof of the Calabi conjecture allows tools of algebraic geometry to be applied to the analytic problem of the existence of a Ricci-flat metric, while duality results from string theory, particularly mirror symmetry, have led to exact results and to insights into new invariants.

This arsenal of mathematical methods has seen heavy use in the construction of realistic vacua of string theory built on compactifications on Calabi-Yau threefolds.  These solutions bear the imprint of the tools used to make them: qualitative properties of the effective theories of Calabi-Yau compactifications differ from those described in \S\ref{sec:power-law}.  Some such properties follow from supersymmetry and the associated nonrenormalization theorems, while others are consequences of particularities of Calabi-Yau geometry itself.  In this section we will summarize recent progress in finding cosmological solutions in Calabi-Yau compactifications.

In a compactification of type IIB string theory on an O3/O7 orientifold $X$ of a Calabi-Yau threefold, the classical superpotential is determined by quantized fluxes $F_3$ and $H_3$, the axiodilaton $\tau$, and the holomorphic $(3,0)$ form of $X$ \cite{Gukov:1999ya}:
\begin{equation}
W_{\text{flux}} = \sqrt{\tfrac{2}{\pi}}\int_X \left(F_3 - \tau H_3\right) \wedge \Omega\,.
\end{equation}
For sufficiently generic choices of the flux quanta, the resulting potential for the complex structure moduli has isolated minima, while the K\"ahler moduli remain unstabilized at leading order in the $g_s$ and $\alpha'$ expansions \cite{Giddings:2001yu}.  The order parameter measuring supersymmetry breaking in these configurations is the expectation value
\begin{equation}
W_0 := \langle W_{\text{flux}} \rangle\,,
\end{equation} where the brackets denote evaluation on the vevs of the complex structure moduli.

The classical flux compactifications of \cite{Giddings:2001yu} do not yield realistic effective theories, because the unstabilized K\"ahler moduli encode instabilities, and are sharply constrained by fifth-force tests and by cosmological limits on moduli.  A natural idea is  to compute quantum corrections order by order in the $g_s$ and $\alpha'$ expansions, and search for isolated local minima of the resulting quantum-corrected potential energy.  This approach of perturbative moduli stabilization may ultimately reveal cosmological solutions, but at present the technical task of computing (rather than modeling, cf.~e.g.~\cite{Berg:2007wt,Cicoli:2021rub}) enough terms is out of reach: see \cite{Berg:2005yu}.\footnote{For a recent proposal to achieve stabilization in the perturbative regime via resummation of leading logarithms, see \cite{Burgess:2022nbx}.}

An alternative strategy \cite{Kachru:2003aw} is to compute \emph{nonperturbative} quantum effects, and then find parameter regimes in which these effects govern the vacuum structure and give rise to isolated vacua.  At first sight this computation sounds even more difficult than determining perturbative corrections.  However, supersymmetry, nonrenormalization theorems, and properties of Calabi-Yau geometry make the task of computing the leading instanton effects a problem in computational topology that turns out to be solvable.

The leading approaches to nonperturbative stabilization are the KKLT scenario \cite{Kachru:2003aw} and the Large Volume Scenario (LVS).  Both rely on Euclidean D3-brane contributions to the superpotential \cite{Witten:1996bn} in order to stabilize the K\"ahler moduli.  The LVS vacuum structure also rests on a well-known correction to the K\"ahler potential at order $\alpha'^3$ \cite{Becker:2002nn}, and, in general, on string loop corrections.

A type IIB flux compactification can contain a supersymmetric AdS$_4$ vacuum of KKLT type only if two conditions are met: there must exist enough rigid divisors supporting Euclidean D3-branes (or gaugino condensates on seven-branes) to stabilize the K\"ahler moduli, and the flux superpotential $W_0$ must be exponentially small in order for the minimum to exist in a parameter regime where the $\alpha'$ expansion is well-controlled.

To check the former condition, one needs to compute the topology of divisors in a Calabi-Yau threefold.  When the number $h^{1,1}$ of K\"ahler moduli is of order five or smaller, this can be accomplished by the algorithm of \cite{Blumenhagen:2010pv}, but vacua have not yet been found in this way.  A tour de force analysis in
\cite{Denef:2005mm} established the existence of enough Euclidean D3-branes in a highly symmetric example with $h^{1,1}=51$, but generalization of this example is an open problem.  More recent computational advances \cite{CYTools,Kim:2021lwo} (see \cite{Braun:2017nhi,Constantin:2018hvl,Brodie:2021zqq} for related techniques) have enabled computation at arbitrary $h^{1,1}$, and enumerating Euclidean D3-brane superpotential terms in Calabi-Yau hypersurfaces is now fast and routine \cite{Demirtas:2021nlu}.  Moreover, the Pfaffian prefactors of such terms are increasingly well-understood \cite{Sen:2021tpp,Demirtas:2021nlu,Alexandrov:2021dyl,Kim:2022jvv}.

The second necessary condition, exponential smallness of $W_0$, involves solving a Diophantine equation for the flux quanta.  It was shown in \cite{Denef:2004cf} (see the review in \cite{Douglas:2006es}) that in the approximation that the fluxes are continuous, and for large enough $h^{2,1}$, small $W_0$ should exist.  However, configurations with small $W_0$ are rare, and even with powerful methods an optimized search is difficult \cite{Cole:2019enn}.

Mirror symmetry provides a solution \cite{Demirtas:2019sip}. One can choose quantized fluxes to make all the terms in $W_{\text{flux}}$ that are polynomial in the complex structure moduli vanish \emph{exactly}.  What remains are corrections that are exponentially small near large complex structure, and if these can be computed, one can arrange to balance two terms against each other in a racetrack.
From the perspective of type IIA string theory on the mirror threefold the corrections in question are worldsheet instanton contributions to the prepotential, while in type IIB they are purely geometric, and can be extracted by computing the periods of $\Omega$ in an integral basis of three-cycles.  The logic for such a computation is well-established \cite{Candelas:1990rm,Hosono:1993qy}, but performing it for $h^{2,1} \gtrsim 5$ has been out of reach of published software.  The algorithms given in the software package {\tt{CYTools}} \cite{CYTools} solve this problem.  With the instanton series in hand, one can find examples where the Gopakumar-Vafa invariants of curves,
and the possible choices of quantized fluxes, combine to yield a racetrack superpotential that in turn leads to exponentially small $W_0$.   Many such examples have been studied \cite{Honma:2021klo,Broeckel:2021uty,Bastian:2021hpc,Carta:2021kpk,Carta:2022oex},
including cases with conifold regions \cite{Demirtas:2020ffz,Alvarez-Garcia:2020pxd} (see also the related works \cite{Marchesano:2021gyv,Grimm:2021ckh}).  Explicit solutions with $W_0$ as small as $10^{-90}$ have been found \cite{Demirtas:2021nlu}.

Combining the above advances in computing Euclidean D3-brane superpotentials and flux superpotentials led to the first explicit, parametrically controlled incarnations of the KKLT scenario \cite{Demirtas:2021nlu,Demirtas:2021ote}.  These vacua are supersymmetric AdS$_4$ solutions, with all moduli stabilized, in which the magnitude of the vacuum energy is exponentially small --- in some cases, smaller than $10^{-123} M_{\text{pl}}^4$.  Such configurations are a step toward specific de Sitter solutions of KKLT type, but several challenges remain.

One issue is that breaking supersymmetry by means of an anti-D3-brane, as in \cite{Kachru:2002gs}, sources corrections to the local field configuration \cite{Bena:2009xk}.
Computation of the backreaction \cite{McGuirk:2009xx,Dymarsky:2011pm}, effective field theory arguments \cite{Michel:2014lva,Polchinski:2015bea}, and analysis of temperature-dependence \cite{Armas:2018rsy} give evidence for metastability, and the constrained supermultiplet formalism provides a simple packaging of the anti-D3-brane action \cite{FKL,Cribiori:2019hod}.
However, the anti-D3-brane is a small perturbation when the Klebanov-Strassler throat region is large, and is supported by a large quantity of D3-brane charge, so one can ask whether throats large enough for metastable anti-D3-branes can be engineered in compact models \cite{Carta:2019rhx}, which necessarily allow only a finite D3-brane charge tadpole \cite{Bena:2020xrh}.
Related tadpole constraints have been studied in \cite{Bena:2021wyr,Gao:2022fdi}.

Furthermore, KKLT vacua with large throat regions have been argued to suffer from singularities caused by large negative D3-brane charge induced on stacks of seven-branes \cite{Gao:2020xqh}.
Evidence from a $\mathcal{N}=2$ model in \cite{Carta:2021lqg} suggests that the seven-branes in question are bound states of exotic branes, whose rearrangement dynamically removes the singularity.  Moreover, the extent of the problem of gluing a throat into a non-singular bulk depends on the typical volume of a Calabi-Yau threefold with $h^{1,1} \gg 1$ K\"ahler moduli.
It was noted in \cite{Demirtas:2021nlu}, following \cite{Demirtas:2018akl}, that at the point in moduli space where the K\"ahler moduli are stabilized by Euclidean D3-branes, the threefold volume scales as a power of $h^{1,1}$, dramatically alleviating the problem of singularities.

A further question is the ten-dimensional field configuration corresponding to a KKLT vacuum.
The interplay between the classical flux energy and the quantum effects of Euclidean D3-branes takes a simple form in the four-dimensional effective theory \cite{Kachru:2003aw}, whereas the effects of anti-D3-branes on fields other than the overall volume are most naturally analyzed in ten-dimensional supergravity (see e.g.~\cite{Baumann:2006th,Baumann:2010sx,Dymarsky:2011pm}).
One can benefit from both perspectives by representing all sources of stress-energy, including four-dimensional quantum effects, in terms of a ten-dimensional solution.

Early work established that the quantum effects of Euclidean D3-branes, or gaugino condensation on D7-branes, lead to a generalized complex geometry
\cite{Koerber:2007xk}, with the gaugino bilinear sourcing fluxes \cite{Baumann:2010sx} and deformations of the metric \cite{Dymarsky:2010mf}.
Constructing a consistent global solution,
following initial steps in
\cite{Koerber:2007xk,Moritz:2017xto}, required further analysis
\cite{Gautason:2018gln,Hamada:2018qef,Gautason:2019jwq,Hamada:2019ack,Carta:2019rhx,Bena:2019mte,Kachru:2019dvo,Randall:2019ent,Hamada:2021ryq}, particularly of four-fermion couplings \cite{Hamada:2018qef,Hamada:2019ack,Kachru:2019dvo,Hamada:2021ryq}
and generalized complex geometry
\cite{Bena:2019mte,Kachru:2019dvo}.
Upon including the gaugino bilinear expectation value, the couplings of D7-brane gauginos to bulk fields couplings source a ten-dimensional solution that corresponds to the vacuum of the four-dimensional effective theory \cite{Hamada:2019ack,Kachru:2019dvo}. The match was made precise in \cite{Kachru:2019dvo} using nonperturbative corrections to the Killing spinor equations, and was derived in a manifestly local manner in \cite{Hamada:2021ryq}.  In summary, the ten-dimensional description meets all consistency requirements and can be applied to study couplings in global models.

A qualitatively important feature
of nonperturbatively-stabilized Calabi-Yau compactifications, such as KKLT vacua, is the impact of nonperturbative couplings on inflationary observables.
Such couplings are essential for computing the vacuum energy, so it is not surprising that they might impact other observables.  Even so, exposing such effects has proved subtle. An early example is D-brane inflation, where leading contributions to the inflaton mass come from inflaton-dependent fluctuation determinants \cite{Kachru:2003sx}, as computed in \cite{Ganor:1996pe,Berg:2004ek,Baumann:2006th}: see \cite{Baumann:2014nda}.

Axion monodromy inflation
is another case
where nonperturbative couplings can affect dynamics around
nonperturbatively-stabilized vacua.
Displacing an axion through one period of monodromy generally leads to the accumulation of one unit of a quantized charge.
In models with moving D-branes, this is often an induced charge on a D-brane, but in other cases it can be a bulk flux.
The stress-energy of this monodromy charge serves as a source term in the ten-dimensional Einstein equations, leading to backreaction on the metric and on other fields. As explained in \S\ref{sec:power-law}, in perturbatively-stabilized models this effect typically leads to computable flattening of the potential over a large inflationary field range.  However, in nonperturbatively-stabilized models the backreaction of monodromy charge can be more problematic, because the actions of Euclidean D-branes depend sensitively on the fields sourced by monodromy charge.
Furthermore, in the subclass of models
where the inflaton is a moving D-brane, the relevant backreaction is that of a localized source in ten dimensions.  The often-employed shortcut of dimensional reduction in the probe approximation, followed by analysis of adjustments in the vevs of fields much lighter than the Kaluza-Klein scale, captures only a subset of the leading backreaction effects, and so can understate the extent of the effect. Indeed, the adjustment of heavy fields during inflation is  phenomenologically relevant in general, and is an example of the UV sensitivity of inflation and its observables \cite{Dong:2010in,Renaux-Petel:2015mga} --- this is both a challenge and an opportunity.

The issue of backreaction was addressed in early examples, indicating that model-building can mitigate the problem in Calabi-Yau realizations of axion monodromy, as shown in
\cite{McAllister:2008hb,Flauger:2009ab,McAllister:2016vzi}.
However, no silver bullet solution is known,
even in F-term models \cite{Marchesano:2014mla,Hebecker:2014eua,Hebecker:2014kva} and Higgs-otic models
\cite{Bielleman:2016olv}: in the latter case, induced charge on a moving D7-brane has a large effect \cite{Kim:2018vgz}.
This general phenomenon could limit embeddings of the relaxion idea \cite{Graham:2015cka} in Calabi-Yau compactifications \cite{McAllister:2016vzi}.

Constraints of this sort can give useful structure to the space of cosmological models descending from string theory:
the small fundamental periods of axions in weakly-coupled limits of string theory make
oscillatory features a natural concomitant of primordial gravitational waves, while backreaction leads to drifts in the frequency and amplitude of the oscillations \cite{Flauger:2009ab, McAllister:2014mpa}.\footnote{The phenomenon of drifting oscillations in axion monodromy also occurs in the examples discussed in \S\ref{sec:power-law} \cite{McAllister:2014mpa}.}

Future work on cosmologies in Calabi-Yau compactifications may be accelerated by improvements in computation, both through computational algebraic geometry, as in \cite{2000math1159E,Blumenhagen:2010pv,Constantin:2018hvl,CYTools}, and through advances in machine learning (see \S\ref{sec:ML}).  Computing all the leading superpotential terms in large classes of flux compactifications is becoming feasible, even in cases with many moduli.
This capability makes it possible to discover new structures and new phenomena in this corner of the landscape.
On the other hand, controlling corrections to the K\"ahler potential remains an important open problem.

\section{Effective field theory and cosmology}

There is a long history of developments in cosmology and in quantum field theory tracing each other. Insights from quantum field theory motivate new approaches to cosmological questions, and cosmology provides a unique arena to study high energy physics in extreme situations. This connection is in turn informed by insights and constraints from ultraviolet completion in string theory. As we recall below, in several cases EFTs were generalized to incorporate phenomena that had been discovered initially in the context of string theory, exposing hidden assumptions in the previous bottom-up theory.

The state of the primordial quantum fields is in general an unknown wave functional of a scalar $\zeta$, a tensor $\gamma$, and additional fields $\chi$.
The likelihood, or probability of the data given the theory, is given by squaring and tracing over the non-observable fields:
\begin{equation}\label{Likgen}
{\cal L}(\zeta(x), \gamma(x) |\{\lambda\})= \int D\chi |\Psi(\zeta(x), \gamma(x), \chi(x); \{\lambda\})|^2.
\end{equation}
Generic inflationary theories generate nontrivial contributions to the power spectrum,
along with non-Gaussian contributions at some level,
with a wide variety of possible functional forms.

There is not yet a systematic understanding of the space of viable states and signatures.  Recent advances very roughly divide into two themes: in one direction, researchers introduce symmetry assumptions enabling streamlined calculations of correlation functions and derivations of theorems organizing their formal properties related to the unitarity and locality of the theory.  In another direction, calculable contributions to the full probability distribution function \eqref{Likgen} beyond perturbative low-point correlators arise in a variety of well-motivated models with more generic interactions; in this same vein of nonperturbative analysis one can address infrared issues in inflation.
In this section we briefly review these threads, along with important related results concerning initial conditions and the robustness of inflation at early times, and IR dynamics at late times.

\subsection{EFT approach to inflation}

A complementary approach to the top-down construction of
ultraviolet realizations of inflation
is to develop an effective description of the dynamics at the level of perturbations around the inflationary background. In~\cite{Creminelli:2006xe,Cheung:2007st}, such a bottom-up ideology was beautifully developed, and goes under the name of the Effective Field Theory of Inflation (EFTofI).
Since observable effects originate at energy scales of order $H$, we may integrate out the UV to write down an EFT that
captures a large class of inflationary models and reveals their universal properties. This universality follows from viewing inflation as a process of spontaneous symmetry breaking, which implies that the effective description of inflation necessarily contains two light degrees of freedom: the scalar Nambu--Goldstone boson for the broken symmetries, and the graviton.
Additional fields can also appear, leading to a wide variety of effects, including non-adiabaticity and a rich space of phenomenological signatures.  Even mass scales $\gg H$ can contribute significantly to observables, as we will discuss further below.  Certain effects of light sectors have been incorporated in the effective description~\cite{Senatore:2010wk,Noumi:2012vr,Lee:2016vti}, some of which can be motivated by spontaneously-broken SUSY~\cite{Baumann:2011nk,Kahn:2015mla,Delacretaz:2016nhw}.
A particularly interesting possibility is that there are degrees of freedom with masses of order the Hubble scale during inflation. These particles cannot be integrated out and modeled by effective interactions of the inflaton, and consequently they would have striking signatures~\cite{Chen:2009zp,Assassi:2012zq,Noumi:2012vr,Arkani-Hamed:2015bza,Lee:2016vti}, providing an interesting probe of high-energy physics. These signatures are challenging to compute with traditional techniques, motivating the bootstrap approaches we discuss in the following.
More generally,
the EFT approach makes it possible to systematically study the space of possible non-Gaussian signatures beyond the simplest slow-roll models of inflation, and has motivated the search for the most symmetric of these shapes in data, which is now standard.  We note that this development illustrates the fruitful collaboration between top-down and bottom-up ideas, as the former helped stimulate the more systematic EFT analysis \cite{Flauger:2009ab,Behbahani:2011it,Senatore:2016aui, Alishahiha:2004eh}.

Interestingly, not all observable effects that are natural from the ultraviolet point of view can be captured by a standard EFT Lagrangian with local interactions of the light fields. As inflation occurs, the inflaton might travel a large distance in the field space, resulting in a significant change of the relevant interactions. One manifestation of this is the flattening effect discussed above \cite{Dong:2010in}. Another  is that additional fields may become light, or lighter, at some moment in time and then become heavy again. Such effects generically lead to some level of `features' in the correlation functions of primordial fluctuations
\cite{Slosar:2019gvt}, often through resonant interactions \cite{ChenEastherLim}.
In the string-theoretic models described above, features can have a particular pattern governed by the discrete shift symmetry of the inflaton~\cite{Flauger:2009ab}. One can incorporate such a symmetry in the EFT as in \cite{Behbahani:2011it}, although drifts in the oscillations are important to capture the string-theoretic phenomenology \cite{Flauger:2014ana}. An important challenge is to understand how to approach apparent breakdowns of the effective description from the bottom up when they occur.  A case of this sort described below in \S\ref{sec:nonadiabatic} has this character and exhibits sensitivity to produced particles with mass $\sim 100 H$ \cite{Flauger:2016idt}.

\subsection{The cosmological bootstrap}

One reason the inflationary period is so mysterious is that we have no direct observational access to this epoch.
Instead, we infer the existence of a period of
inflation by measuring correlations on its late-time boundary, where the universe reheats. This suggests a different approach to
understanding the physics of inflation, where we directly construct boundary correlations, rather than following inflationary time evolution. This approach
harmonizes
with EFT approaches to cosmological correlations (e.g.,~\cite{Pajer:2020wxk,Bonifacio:2021azc}), and leverages developments in the study of scattering amplitudes, the conformal bootstrap, and holography to import insights into the cosmological setting.
Work in this area has mostly focused on the simplified setting where de Sitter symmetry is either exact, or weakly broken, at the level of the background, and gravitational fluctuations are treated only perturbatively.
The high degree of symmetry allows for analytic control, but it is important to understand how to move beyond such simplifications. In the following we briefly review some recent progress and future directions.\footnote{For a more comprehensive discussion, see the Snowmass white paper on the Cosmological Bootstrap~\cite{Snowmass2021:CosmoBoot}.}

In perturbation theory, cosmological correlation functions, at least given the setup and simplifications noted above,
are in large part controlled by their singularity structure~\cite{Maldacena:2011nz,Raju:2012zr,Arkani-Hamed:2015bza,Arkani-Hamed:2017fdk,Arkani-Hamed:2018kmz,Arkani-Hamed:2018bjr,Benincasa:2019vqr,Benincasa:2018ssx,Baumann:2020dch}, which encodes the locality of bulk interactions.
One can then approach the construction of cosmological correlation functions from a viewpoint that is spiritually similar to the flat space amplitudes program~\cite{elvang_huang_2015}. The challenge is to systematically extend correlators away from their singularities, which typically requires some additional input.  One approach is to utilize de Sitter symmetry, which implies differential equations solved by cosmological correlators~\cite{Maldacena:2011nz,Bzowski:2013sza,Arkani-Hamed:2015bza,Arkani-Hamed:2018kmz,Baumann:2019oyu,Baumann:2020dch,Bzowski:2019kwd}. A complementary tack is to utilize constraints from the unitarity of bulk time evolution~\cite{Goodhew:2020hob,Cespedes:2020xqq}, which can be systematized into perturbative cutting rules for the cosmological wavefunction, similar to flat space Cutkosky rules~\cite{Meltzer:2020qbr,Melville:2021lst,Goodhew:2021oqg,Benincasa:2020aoj,Baumann:2021fxj}. These relations take the schematic form
\begin{equation}
    \psi_n(\{k_a\})+\psi_n^*(\{-k_a\}) = -\sum_{\rm cuts}\,\psi_n\,,
    \label{eq:unitarity1}
\end{equation}
where $\psi_n$ is an $n$-point cosmological wavefunction coefficient, and $\{k_a\}$ stands for the set of magnitudes of momenta that the wavefunction depends on.\footnote{See~\cite{Goodhew:2020hob} for the precise analytic continuation of $k_a\to-k_a$.} In many cases of phenomenological interest involving massless particles, information about these cuts is sufficient to reconstruct wavefunction coefficients in a de Sitter background (but allowing interaction vertices to violate the symmetries, as in the EFT of inflation)~\cite{Jazayeri:2021fvk,Bonifacio:2021azc,Cabass:2021fnw,Hillman:2021bnk,Meltzer:2021zin,Baumann:2021fxj}. In cases involving massive particles, one must instead use the full differential equations that the correlators satisfy. Significant progress has been made in characterizing the possible signatures of these particles~\cite{Chen:2009zp,Arkani-Hamed:2015bza,Arkani-Hamed:2018kmz,Sleight:2019hfp,Baumann:2019oyu}, though the signals are expected to be weak where the calculations are under control, so it is important to further develop these techniques.

A complementary approach that utilizes de Sitter symmetry in a more direct way is based on the spectral decomposition of cosmological correlators with respect to conformal partial waves~\cite{DiPietro:2021sjt,Hogervorst:2021uvp}. This representation  highlights different analytic properties of the observables, and establishes a different condition following from unitarity of the cosmological evolution. For example, for the four-point function of identical fields, unitarity implies positivity of densities $\rho_J(\nu)$, which are defined as
\be
\langle{\phi(x_1)\phi(x_2)\phi(x_3)\phi(x_4)}\rangle=\sum_J \int_0^{\infty}d\nu \rho_J(\nu){\mathcal{F}}_{J,\Delta}\,,
\label{eq:unitarity2}
\ee
where ${\mathcal{F}}_{J,\Delta}$ is the conformal partial wave of spin $J$ and dimension $\Delta=d/2+i\nu$. It is natural to expect that~\eqref{eq:unitarity1} and~\eqref{eq:unitarity2} are both manifestations of the same underlying unitarity of physics in de Sitter space, but their relation has not yet been fully understood. Making such a connection is an important direction for the future.

A useful practical tool for the calculation of correlators is the analytic continuation between in-in dS observables and Euclidean AdS boundary correlators in a theory with the doubled set of fields~\cite{DiPietro:2021sjt,Sleight:2020obc,Sleight:2019mgd,Sleight:2019hfp,Sleight:2021plv}. This continuation makes it possible to exploit many powerful AdS techniques for cosmology.

A  goal for the future is to sharpen the connections to ultraviolet physics from the bottom-up EFT perspective. One facet of this connection manifests through dispersion relations obeyed by correlation functions, and it will be very interesting to elucidate the structure of such relations in the cosmological setting. This will require a more refined understanding of the analytic structure of correlators, and of the constraints of unitarity. It is also important to further develop nonperturbative techniques in the cosmological setting, in order to place sharp bounds on the space of inflationary models and their particle content, in a similar spirit to the conformal bootstrap in the AdS context. We expect that insights from ultraviolet completions will play a crucial role in these developments, leading to a fruitful interplay between UV and IR.

\subsection{Initial conditions classically}\label{sec:classical-initial}

Once inflation begins, it is extremely robust, but it is often stated that inflation has an `initial conditions problem' and only begins if the universe is homogeneous and isotropic over several Hubble scales~\cite{Goldwirth:1989pr}.
This problem is absent in models of chaotic inflation in which the inflationary period is described by general relativity coupled to a scalar field and begins with nearly-Planckian energy density or equivalent~\cite{Linde:1985ub,Linde:1990flp,Linde:2014nna}.
Observations of the CMB disfavor the simplest version of this solution, but certain modified models are still allowed~\cite{Mukhanov:2014uwa}. In addition, in the context of a weakly coupled UV completion of general relativity, additional scales appear below the Planck scale, so the effective description in terms of general relativity coupled to a scalar field only becomes valid well below the Planck scale. As a consequence, both recent observations and theoretical considerations motivate a closer look at inflation's robustness to inhomogeneities.

To the extent that the potential energy density of the scalar field is well-modeled by a cosmological constant, de Sitter no-hair theorems carry over to inflation. These theorems come in two flavors, one for homogeneous cosmologies and one for inhomogeneous cosmologies. In the homogeneous case, the possible cosmologies are typically referred to according to the Bianchi classification for the algebra of their Killing vectors. Except for Bianchi IX,  all homogeneous anisotropic cosmologies with positive cosmological constant asymptote to de Sitter space~\cite{Wald:1983ky}. In the inhomogeneous case, provided the weak energy condition holds, it can be shown that global recollapse can only occur if the three-dimensional Ricci scalar is positive everywhere, which is topologically impossible for most 3-manifolds~\cite{1985MNRAS.216..395B,Kleban:2016sqm}. Moreover, under the same assumptions it can be shown that there is always an open neighborhood that expands at least as fast as de Sitter space~\cite{Kleban:2016sqm}. More recently, it has also become possible to show a stronger statement using mean curvature flow, namely that under additional assumptions the spacetime asymptotically becomes indistinguishable from de Sitter space on arbitrarily large regions of spacetime~\cite{Creminelli:2020zvc}, and that this still holds if the cosmological constant is replaced by a scalar field with a potential, provided the latter is bounded both above and below by positive cosmological constants whose ratio is between unity and $3/2$~\cite{Wang:2021hzv}.

In general, the field will explore regions of the potential that do not support inflation.  Studies of this regime currently rely on numerical simulations of general relativity coupled to a single scalar field. This line of research has a long history and includes work in 1+1 dimensions~\cite{Albrecht:1986pi,Kurki-Suonio:1987mrt,Goldwirth:1989pr} and 3+1 dimensions~\cite{Laguna:1991zs,Kurki-Suonio:1993lzy}. The advent of modern numerical general relativity codes and more powerful computers has provided motivation to revisit this question. Importantly, it is now possible to follow the time evolution in a regime in which black hole formation plays a significant role
\cite{East:2015ggf,Clough:2016ymm}. The new simulations have shown that inflation is more robust than previously thought and can begin even for classes of initial conditions for which the gradient energy density exceeds the potential energy density by orders of magnitude, both for plateau models~\cite{East:2015ggf} and for models of chaotic inflation in which the energy density is well below the Planck scale~\cite{Clough:2016ymm}. Additional classes of initial conditions and shapes of the potential have been considered~\cite{Clough:2017efm,Aurrekoetxea:2019fhr,Joana:2020rxm}, leaving the conclusions unchanged but sharpening the picture. Models with convex potentials are found to be more robust to inhomogeneities than those with concave potentials, and concave potentials that vary on super-Planckian scales are significantly more robust than those that vary on sub-Planckian scales~\cite{Aurrekoetxea:2019fhr}.  Notably, the classes of models that appear robust to large initial inhomogeneities are the classes of models that predict gravitational wave signals that are large enough to be detected with upcoming CMB experiments. Therefore, in the absence of a detection of primordial gravitational waves, additional dynamics (such as an additional scalar field that drives an earlier period of inflation, or a tunneling process) may be required to prepare initial conditions appropriate for inflation in the remaining class of models with sub-Planckian characteristic scales.

\subsection{Initial conditions semiclassically}

Above we reviewed the beginning of inflation from a large set of initial conditions at the classical level. Ultimately, one would like to develop a theory that is capable of explaining the initial conditions for the universe at the full quantum level. While this problem is notoriously hard, there exists an effective semiclassical approach that is conjectured to incorporate quantum gravity effects relevant in the early universe without using an explicit UV theory. The original proposal \cite{hartle1983wave}, now 40 years old,
postulates that at the early time the metric should be analytically continued to become Euclidean and smooth, thus avoiding the curvature singularity that would be present if one evolves the universe back in time and stays in the Lorentzian signature.

This idea, referred to as the no-boundary proposal, has an important drawback in inflation. Interpreted in the most straightforward way, it predicts the probability for the initial value of the inflaton field, $\phi_0$, which goes as $p(\phi_0)\sim \exp{1/V(\phi_0)}$, thus peaking exponentially at the bottom of the inflaton potential. Thus the no-boundary proposal predicts almost no inflation, in contrast with observations. There have been multiple proposals to fix this problem, while maintaining the spirit of the idea, including volume weighting \cite{Hartle:2007gi}, which changes the probability to $p(\phi_0)\sim \exp{1/V(\phi_0)+3 N_e}$. In turn the tunneling initial state proposal (see \cite{Vilenkin:2002ev,Linde:1998gs} for reviews) suggests that the probability is $p(\phi_0)\sim \exp{-1/V(\phi_0)}$, thus predicting the start of the classical inflationary regime at the top of the inflaton potential.

The recent revival of interest in the Euclidean Quantum Gravity
(EQG) approach to cosmology is motivated by the success of related ideas in understanding some of the salient features of the black hole evaporation process \cite{Almheiri:2020cfm}. In this case holography combined with the quantum information understanding of the boundary theory allows one to confirm the EQG calculations, which in certain cases include significant contributions from non-trivial topologies.
These insights suggest that non-trivial topologies should also be incorporated in the cosmological no-boundary proposal. These include space-time connections between the bra and ket of the wave function of the universe \cite{Chen:2020tes}, as well as the appearance of `islands' in the calculation of entropy in various regions of the inflating universe: see e.g.~\cite{Chen:2020tes,Hartman:2020khs,Bousso:2022gth,Langhoff:2021uct,Aguilar-Gutierrez:2021bns, Sybesma:2020fxg,Aalsma:2021bit,Kames-King:2021etp}.

So far, explicit computation of these new contributions to the gravitational path integral has been possible only in simplified models of cosmology, mostly in lower dimensions. Whether analogous effects are present in realistic models, and if they can be observable, remains an intriguing direction for future research.

More generally, quantum gravity admits a multitude of states even in a closed system.  Indeed, in the previous section we elucidated the robustness of inflation in the presence of strong initial fluctuations.  The two dimensional quantum gravity theory on the closed string worldsheet is a clean example illustrating the multiplicity of states, with production of strings (two-dimensional universes) arising in generic target spacetimes.  This can be considered with particular worldsheet models corresponding to $D=2$ de Sitter and inflationary physics, as in \cite{CarneirodaCunha:2003mxy}; this would be interesting to connect to the more recent developments in this section.

\subsubsection{Spacetime singularities}

The UV completion of gravity is required to understand spacetime singularities, including dynamical topology-changing processes.  A zoo of timelike singularities and some spacelike singularities admit known string-theoretic resolutions, involving the condensation of wrapped branes or wound strings.  Some are perturbative but stringy, while others are nonperturbative. In the cosmological context, an early review can be found in \cite{McAllister:2007bg}.
Related to the study of spacelike singularities is the additional question of alternatives to inflation  such as bouncing or cyclic cosmologies, which are more difficult to control.

\subsection{Nonperturbative physics of inflationary EFT}

In this subsection, we highlight recent advances in understanding nonperturbative physics of inflationary observables and its interplay with UV physics.  In these studies, one moves beyond low-point correlators in highly symmetric models in order to understand more general aspects of \eqref{Likgen}.  Recent highlights include control of IR effects, non-adiabatic effects with signal/noise beyond low-point correlators, and more general contributions to the tails of the distribution.

\subsubsection{Strong IR effects}

Enhancement of IR effects for light fields has long been an intriguing issue in inflationary cosmology. In certain cases perturbation theory around the free-field vacuum breaks down at long distances and times, making it possible to question the stability of the corresponding theories \cite{Polyakov:2012uc}. An approach to deal with these IR effects has been known  for a long time \cite{Starobinsky:1984}, but has not been developed systematically. For light scalars decoupled from gravity this was rectified in \cite{Gorbenko:2019rza}, and confirmed and extended using related but different techniques in \cite{Mirbabayi:2019qtx,Cohen:2020php,Mirbabayi:2020vyt,Baumgart:2020oby,Cohen:2021fzf,Cohen:2021jbo}. The main conclusion is that strong IR effects are physical, can be controlled and calculated, and can lead to nonperturbative effects with observable consequences. One such method, called Soft dS Effective
Theory \cite{Cohen:2021jbo}, was recently applied to inflation, i.e.~dynamical gravity was also incorporated. The conclusion of this paper is that strong IR effects can have interesting implications for eternal inflation and lead to a breakdown in the EFT of perturbations if significant (but observationally viable) non-Gaussianities are also present in the model.

\subsubsection{Non-adiabatic effects}\label{sec:nonadiabatic}

Inflation and its observational constraints are compatible with a wide range of interactions, including direct, non-derivative interactions of the inflaton with other degrees of freedom, generically leading to non-adiabatic effects \cite{Chung:1999ve,Romano:2008rr}.  Such interactions arise generically in top-down settings  \cite{Silverstein:2017zfk}\ and may follow a regular \cite{Green:2009ds} or random \cite{Garcia:2020mwi}\ pattern.
Recent works \cite{Flauger:2016idt,Munchmeyer:2019wlh} established observational sensitivity to production of particles with masses always greater than the Hubble scale during inflation.  (We already saw above in \S\ref{sec:string-inflation} at the level of the background model that massive degrees of freedom adjust in a way that is energetically favorable, affecting predictions for the tilt and tensor to scalar ratio even with masses much greater than the Hubble scale.) This example is notable for two reasons: (i) it goes beyond the naive EFT description, requiring inclusion of these massive degrees of freedom given the precision of modern data, and (ii) it exhibits greater signal/noise in higher $n$-point functions with $n>3$, with the resummed optimal estimator for the simplest version of the signal derived in \cite{Munchmeyer:2019wlh}.

This raises the more general question of the information content of high-point scalar correlators and the tails of the primordial non-Gaussian probability distribution.
We summarize recent developments in this area next.

\subsubsection{Tails of the primordial non-Gaussianity}

A key research direction is to derive the information content in \eqref{Likgen}\ that is encoded beyond low-point correlation functions.  Advances in this direction via analytic and numerical methods going back to \cite{Bond:2009xx} include calculable models with heavy tails at the multifield \cite{Panagopoulos:2019ail, Chen:2018uul,Panagopoulos:2020sxp} and single-field \cite{Celoria:2021vjw} levels.
One example with a very heavy tail  explored in \cite{Panagopoulos:2019ail,Panagopoulos:2020sxp} derives from the kinetic term for inflation on a multidimensional field space with hyperbolic geometry, a possibility with theoretical underpinnings \cite{Brown:2017osf, Carrasco:2015iij}.  This model involves an interaction of the form $\dot\phi^2 \text{exp}(\chi/M_*)$: expanding this coupling around the background homogeneous $\dot\phi$ evolution yields a term in the Hamiltonian proportional to the field momentum for $\phi$, which generates a translation in its field space, leading to an explicit nonperturbative expression for the non-Gaussian pdf with a heavy tail $\propto\,\text{exp}\left\{-(\text{log}(\delta \phi^2))^2 M_*^2/H^2\right\}$.
Another class of examples concerns propagation of the inflaton on a landscape with saddle points, leading to strong effects of the temporarily tachyonic transverse scalars.  Simulations yield interesting phenomenology in ongoing work on this case \cite{BBGM} (see also \cite{Dias:2017gva}).

In general, these examples motivate a more systematic study of the full pdf and its optimal estimation in data.  This is a compelling theoretical challenge on two fronts, involving analysis of the nonperturbative early universe quantum field theory as well as analysis of LSS, perhaps itself requiring a nonperturbative treatment.  In the examples covered in \cite{Munchmeyer:2019wlh}, a position space description of the primordial signal is natural, and this may combine with the need to measure high-point correlators and/or to work at the map level for the foreground in any case \cite{Baumann:2021ykm}.

\section{Cosmological holography}\label{sec:holography}

Cosmology ultimately demands both a UV and an IR completion, with quantum gravity operating in both regimes.  In particular, the global spacetime(s) of interest that contain long-lived regions of metastable de Sitter
do not fully decouple from gravity at finite times, as a result of their finite size.  The metastability --- a property most clearly derived via string theory \cite{Dine:1985kv,Giddings:2003zw} --- does enable asymptotic decoupling of gravity and the development of infinite horizon entropy \cite{Bousso_2002,Dong:2011uf}.    Even in the long-lived de Sitter phase, in addition to the EQG developments summarized above, various tools such as
solvable deformations that enable construction of emergent patches of spacetime, as well as other developments in low-dimensional gravity and matrix models, have led to significant recent progress in this area.

In this section, we summarize this renewed traction on these old problems, which opens many novel research directions.  A nice review of early developments in this area can be found in \cite{Anninos:2012qw}; see also  \cite{Silverstein:2016ggb} for some intermediate developments.  Pre-existing approaches include dS/CFT \cite{Strominger:2001gp,Anninos:2011ui}, dS/dS \cite{Alishahiha:2004md,Dong:2010pm} and FRW/FRW \cite{Dong:2011uf}, FRW/CFT \cite{Freivogel:2006xu} and various types of matrix models \cite{Banks:2018ypk,Anninos:2011af} (see also \cite{Susskind:2021dfc,Shaghoulian:2021cef,Susskind:2021esx} for more recent discussions), as well as discussions of embedding dS (or a different FRW solution) in AdS/CFT \cite{Anninos:2017hhn, Cooper:2018cmb} and uplifting AdS/CFT to cosmology via explicit gravity-side ingredients \cite{Dong:2010pm, DeLuca:2021pej}.

Gibbons and Hawking suggested an entropic interpretation of the de Sitter horizon area \cite{Gibbons:1977mu}.  This classical entropy $S_{GH}=\text{Area}/4 G_{N}$
has been generalized to higher loop order in \cite{Anninos:2020hfj}.  Together these results suggest that as with black holes, there is a large set of microstates for which the cosmic horizon represents a
coarse-grained
description.

In the black hole case,
the corresponding microstates
lie in a small band of energy levels with small level spacings of order $\text{exp}(-S)$. For a three-dimensional bulk, modular bootstrap methods enable direct analysis of this band of energies on the dual two-dimensional side of the correspondence  \cite{Hartman_2014,Mukhametzhanov:2019pzy}, showing that the Cardy formula for their entropy --- which matches the $\text{Area}/4 G_N$ prediction --- extends down to the Hawking-Page transition.

In black hole physics, the count of microstates
of BPS black holes \cite{Strominger:1996sh} from brane degrees of freedom was enabled by extended supersymmetry. Recently a new type of controllable deformation enabling the tracking of energy levels and state counts has emerged \cite{Zamolodchikov:2004ce,Smirnov:2016lqw,Cavaglia:2016oda,Dubovsky:2012wk,McGough:2016lol,Kraus:2018xrn,Gorbenko:2018oov,Lewkowycz:2019xse}, and is based on integrability ideas rather than supersymmetry. These integrable deformations --- of in general {\it non-integrable} `seed' theories --- address the leading count of states for the cosmic (de Sitter) horizon in appropriate examples \cite{Coleman:2021nor,Shyam:2021ciy}.

In this vein,  \cite{Shyam:2021ciy,Coleman:2021nor} derives the $\mathrm{dS}_3$ geometry and the Gibbons-Hawking entropy --- including a 1-loop  correction in \cite{Anninos:2020hfj} --- from a solvable deformation of a CFT \cite{Gorbenko:2018oov,Lewkowycz:2019xse} with a sparse light spectrum.  The seed CFT counts the $\Delta=c/6 +{\cal O}(1)$ BTZ entropy as in \cite{Hartman_2014,Mukhametzhanov:2019pzy}, and this state count continues to hold in the $T\bar T$ deformed theory, whose gravity-side description  consists of a bounded patch of the corresponding BTZ black hole geometry \cite{McGough:2016lol}.
Deforming this seed CFT on a cylinder of circumference $L$ as $\partial_\lambda \log(Z)\sim \int T\bar T $ until $y=\lambda/L^2$ reaches $3/c\pi^2$, and then by $\partial_\lambda \log(Z)\sim \int T\bar T -2/\lambda^2$ to a final value $y$, yields a dressed energy formula for the $\Delta=c/6+{\cal O}(1)$ states matching precisely the Brown-York energy of the
cosmic horizon region of the observer patch,
bounded by a cylinder of size $L=\sqrt{\lambda/y}$.  At the same time, the count of states matches the Gibbons-Hawking entropy, including the logarithmic correction computed in \cite{Anninos:2020hfj}.

In more detail, the $T\bar T+\Lambda_2$ deformation for a theory on a cylinder of size $L$ \cite{Gorbenko:2018oov,Lewkowycz:2019xse,Coleman:2021nor} yields a differential equation for the dimensionless energies ${\cal E}=E L$
\begin{equation}\label{eq:Energy-diffeq}
\pi y \mc E(y) \mc E'(y)- \mc E'(y)+ \frac{\pi}{2} \mc E(y)^2=  \frac{1-\eta }{2 \pi  y^2}+2\pi^3 J^2\,,
\end{equation}
where $\eta=1$ for the pure $T\bar T$ part of the trajectory and $\eta =-1$ for the $T\bar T+\Lambda_2$ part.  This has general solution
\be\label{eqn:S1-En-gen}
\mc E(y) = \frac{1}{\pi y} \left(1 \pm \sqrt{\eta-4 C_1 y+4 \pi^4 J^2 y^2} \right)\,,
\ee
whose branch and constants are determined by the boundary conditions along the piecewise trajectories in a given definition of the deformed theory. In the example just summarized, the matching between the $\eta=1$ and $\eta=-1$ parts of the trajectory occurs when the square root vanishes.  Holographically, the boundary of the patch of $\mathrm{(A)dS}_3$ skirts the horizon of a BTZ $\Delta \simeq c/6$ black hole in the AdS case, and the cosmic horizon in the dS case; such near-horizon regions are indistinguishable and the deformed energy formulas match since $\eta$ appears inside the (here vanishing) square root.\footnote{In a complete top-down model, such as uplifts of AdS to dS  \cite{Dong:2010pm, DeLuca:2021pej}, the internal dimensions change.  This fits with the indistinguishability of the systems when the boundary in the external dimensions skirts the  horizon.   For example,  in a canonical ensemble the high temperature near the horizon enables the system to explore all internal configurations consistent with the horizon, including both types of internal compactifications related by internal topology-changing processes.} These energy levels dominate the real spectrum of the deformed theory, and their count of states $S\sim A/4 G_N - 3 \log (A/4 G_N)$ remains
constant through the entire integrable deformation \cite{Coleman:2021nor}.   The bounded patch thus formulated may be viewed as a complete system (dS conditioned on the presence of the boundary, perhaps viewed analogously to conditioning on the existence of an observer) or as a building block for the global system \cite{Coleman:2021nor}.

The deformed theory is a specific matrix theory
that
precisely captures the pure-gravity features --- the geometry, Brown-York energy and entropy --- that dominate at $c\gg 1$, but does not capture the model-dependent details of local bulk physics, as stressed in the AdS case in \cite{Kraus:2018xrn}. The naturally light bulk degrees of freedom are related to gauge fields and may require incorporating other currents into the deformation (see e.g.~\cite{Jiang:2019epa} for a review.)

This set of results consolidates progress on many fronts by a wide variety of researchers, as is often the case for new dualities --- in this case ranging from integrability theory to quantum gravity. They raise numerous directions for future work, not least the generalization to bulk $\mathrm{dS}_4$, perhaps leveraging dimension-independent simplifications of the holographic RG \cite{Dong:2012afa}.

Let us comment briefly on the relation to other approaches.  The static patch Hamiltonian is the modular Hamiltonian of the dS/dS patch, where the entropy just computed corresponds to an entanglement entropy between the two sectors, confirming the interpretation proposed in \cite{Dong:2018cuv}.  Many works stressed the role of matrix models, e.g.~\cite{Banks:2018ypk,Anninos:2011af,Susskind:2021dfc,Susskind:2021esx}, with the deformed theory providing an explicit class of examples.  Calculable recent models with interesting properties include \cite{Anninos:2021ihe,hikida2021holography}.  The particular matrix theory obtained by the integrable deformations is fully nonlinear in the original seed CFT variables, suggesting connections to complexity \cite{Susskind:2021esx}, which is conjecturally related to reconstruction of the region behind the horizon.
Another approach going back to \cite{Freivogel:2005qh}\ is to embed cosmological physics in AdS/CFT \cite{Anninos:2017hhn,Gross:2019ach,Cooper:2018cmb}.  In that case, the theory is UV complete although there are uncertainties in the dictionary extracting the cosmology from the dual theory.  Related works have used probes of microscopic physics such as quantum chaos, which has been fruitful in the context of AdS/CFT, to constrain the holographic dual to de Sitter space \cite{Aalsma:2020aib,Aalsma:2021kle,Anninos:2018svg}.

Last but not least, a  connection between AdS/CFT and de Sitter quantum gravity arises from string-theoretic de Sitter models obtained by explicit uplifts of AdS/CFT systems \cite{Dong:2010pm,DeLuca:2021pej}.  As a start, to connect this
relationship
to the $T\bar T+\Lambda_2$ (or more general $T^2+\Lambda_d$ \cite{Hartman:2018tkw}) approach, one can determine the conditions for a patch boundary in string/M-theory by generalizing the Ho\v{r}ava-Witten analysis of \cite{Horava:1996ma} and determining the implications of the considerations of \cite{Andrade:2015gja,Witten:2018lgb}.
This approach aims at a top-down realization of the interpolation between a radially cut-off AdS black hole and the bounded cosmic horizon patch of de Sitter space.  Recall from above that the $T^2$ and $T^2+\Lambda_d$ trajectories match as the wall approaches the horizon, where in the external dimensions one cannot tell the difference geometrically between the near-horizon region of a black hole and the de Sitter cosmic horizon of the same radius, and the dressed energy formula is continuous along the combined deformation.  In a UV complete model
the internal dimensions are drastically different, e.g.~being a sphere or hyperbolic space for
anti-de Sitter or de Sitter, respectively.  But for the bounding wall skirting the horizon, the diverging temperature forces the full system to sample internal configurations consistent with the horizon geometry.  In that sense, one cannot tell the difference between the two, even including the details of the UV completion at the matching point, so the deformation extends naturally to the full theory.
This interpretation relies on the top-down consistency of the bounded domain and the existence of topology-changing processes
analogous to those in \cite{Adams:2005rb}, both features amenable to further study.\footnote{Investigations of this question by some of the authors of \cite{Coleman:2021nor}\ and \cite{DeLuca:2021pej} are in progress.}

In addition to the `local' approach to cosmological holography, centered around an observer internal to the cosmological space, there is also the `global' approach, in which the holographic theory is formulated on the future spacelike boundary of a de Sitter or inflationary spacetime, and bulk time evolution is a fully emergent phenomenon. As was emphasized in \cite{Maldacena:2002vr}, a crucial step in this approach is to go from the wavefunction of the universe, which can be described by a holographic CFT via dS/CFT, to the correlation functions. Since this step involves a path integral over boundary metrics, it requires the use of tools that go beyond the usual holography. Recent progress in this direction includes, in particular, the study of higher-spin theories in dS, dual to the free $O(N)$ model on the boundary \cite{Anninos:2017eib}.  Recent developments in this interesting direction include \cite{Hikida:2021ese}.  Moreover, Euclidean  calculations starting from global low-dimensional de Sitter such as \cite{Anninos:2021eit, Anninos:2020hfj}\ yield formulas expressible as a sum over microstate contributions.  These in turn may connect to lower-dimensional versions of the $T\bar T+\Lambda$ deformation generalizing \cite{Gross:2019ach} as in \cite{Coleman:2021nor} to yield a Lorentzian microstate count as reviewed for the bulk three-dimensional case above.  Indeed, interesting works in this direction continue to emerge (e.g.~\cite{Svesko:2022txo}).

\section{Observational tests}

\subsection{Tests of inflation}

Many of the recent developments we have highlighted will be tested through cosmological observations. Most prominently, over the next decade precision measurements of the cosmic microwave background polarization both from the ground~\cite{CMB-S4:2016ple} and from space~\cite{LiteBIRD:2022cnt} will begin to cross critical thresholds in the search for primordial gravitational waves. These measurements will, for example, conclusively test the class of models of axion inflation whose potentials are flattened by backreaction relative to a quadratic potential. In this class of models the potential during the inflationary period is well-approximated by a monomial $V(\phi)\approx \mu^{4-2p}\phi^{2p}$, which predicts a spectral index and tensor-to-scalar ratio given by
\begin{equation}
n_{\rm s}=-\frac{p+1}{N_\star}\qquad\text{and}\qquad r = \frac{8p}{N_\star}\,,
\end{equation}
where $N_\star$ measures the number of e-folds (counted from the end of inflation) at which the pivot scale exits the horizon. The numerical value of $N_\star$  depends on the model, the (unknown) physics of reheating, and the choice of pivot scale. If reheating is highly efficient, a typical value is $N_\star\approx60$. If reheating is delayed, $N_\star$ takes a lower value, with the range often take to be $N_\star\approx50-60$, although lower values are possible. For plausible values of $p$, the models predict $r> 0.01$. At face value this class of models already
appears nearly excluded
at the single-field level~\cite{BICEP:2021xfz}, but it should be kept in mind that this conclusion is predominantly based on constraints on the scalar spectral index $n_{\rm s}$, and the theoretical prediction may be modified by the presence of additional degrees of freedom~\cite{Wenren:2014cga}, or may have to be reinterpreted~\cite{DAmico:2021vka,DAmico:2021zdd}.  We note that as discussed above in \S\ref{sec:string-inflation}, since axions arise from the rich topology of the internal dimensions of string theory --- topology that plays an important role in facilitating the near-cancellation of the cosmological constant \cite{Bousso:2000xa} --- the multifield case is a reasonable expectation and is important to test.

More generally, the threshold around $r\simeq 0.01$ is of interest because the distance traversed by the inflaton during the inflationary period is bounded by the tensor-to-scalar ratio~\cite{Lyth:1996im}
\begin{equation}
\frac{\Delta\phi}{M_{\text{pl}}}\gtrsim \left(\frac{r}{8}\right)^{1/2}N_\star\gtrsim\left(\frac{r}{0.01}\right)^{1/2}\,,
\end{equation}
where we have used a conservative lower limit $N_\star\gtrsim 30$ for the last inequality. So a detection at or above this threshold implies a super-Planckian excursion in field space. In any model of inflation in a UV complete description of gravity, like string theory, there are states at sub-Planckian energies that interact with the inflaton and lead to structure in the inflaton potential on sub-Planckian scales, unless these interactions are forbidden by symmetries. So in the context of single-field models a detection of primordial gravitational waves with $r\gtrsim 0.01$ would provide strong support for the existence of an effective shift symmetry that protects the inflaton potential.

The absence of a detection of primordial gravitational waves with a tensor-to-scalar ratio at or above $r\simeq 0.01$ does not imply that the inflaton traveled a sub-Planckian distance. However, over the next decade CMB polarization measurements will cross another critical threshold around $r\simeq 0.001$ that provides related information about the structure of the inflationary potential rather than the field displacement. The monomial models naturally explain the observed spectral index in the sense that the functional form of the spectral index is $n_{\rm s}(N)=-(p+1)/N$. There is a second class of single-field models for which this is the case: hilltop and plateau models. To give concrete examples, this class contains Starobinsky's $R^2$ inflation~\cite{Starobinsky:1980te}, models in which the Higgs boson is the inflaton~\cite{Bezrukov:2007ep,Ballesteros:2016xej}, fibre inflation~\cite{Cicoli:2008gp}, $\alpha$-attractors~\cite{Kallosh:2013yoa,Kallosh:2014rga,Carrasco:2015pla},  and Poincar\'e disk models~\cite{Ferrara:2016fwe,Kallosh:2017ced}.
A key quantity in these models is the scale in field space over which the potential departs from its value on the hilltop or plateau: this is known as the
characteristic scale~\cite{CMB-S4:2016ple}.
The absence of a detection at the threshold around $r\simeq 0.001$ would exclude all models in this class with a characteristic scale that exceeds the Planck scale.

In addition, the same CMB experiments will provide improved constraints on the statistical properties of the primordial density perturbations. Constraints on the spectral index of density perturbations, its running, and on the amplitudes of the traditional shapes of bispectra denoted by $f_{NL}$ will improve by a factor of two to three, significantly reducing the traditional model space.
Three-dimensional surveys, most notably of LSS,
at face value allow access to significantly more Fourier modes. However, extracting information about the primordial universe and in particular about departures from Gaussianity has proven challenging because clustering of matter is itself nonlinear, because surveys observe the galaxy distribution rather than the matter field, because observations take place in redshift space, and so on.  Nonetheless, recent theoretical developments on an effective field theory of large scale structure (EFTofLSS) have matured enough for a first proof of principle constraint on primordial non-Gaussianity~\cite{Cabass:2022wjy,DAmico:2022gki} from BOSS DR 12 data~\cite{BOSS:2016wmc}. As these tools further mature and data from larger surveys become available, constraints from LSS surveys should become competitive with those from CMB observations, and ultimately improve on them. For additional discussion of observational constraints on inflation both in a broader context and from a bottom-up perspective see~\cite{Snowmass2021:DataCosmo} and~\cite{Snowmass2021:inflation}, respectively.

The string-inspired effects we have described  have already motivated~\cite{Silverstein:2017zfk,Baumann:2014nda} and continue to motivate additional analyses to make full use of the data, for example searches for imprints of non-adiabatic effects such as the production of particles or strings during inflation~\cite{Flauger:2016idt}, and more generally observables beyond low order $n$-point functions to fully characterize departures from a Gaussian probability distribution.  The combined challenge of characterizing such effects theoretically and testing them in LSS provides a key goal for theoretical cosmology research in the near and medium term.

A number of other important empirical constraints affect theoretical cosmology, ranging from bounds on axion masses, as discussed in the next subsection,
to constraints on low-energy supersymmetry. Prospects for additional types of tests that could be important, such as new ideas for cosmological gravitational wave sources and reheating effects \cite{Amin:2014eta} deserve continued study.

\subsection{Limits on axions}

As explained in \S\ref{sec:string-inflation}, the effective field theories descending from string theory that are relevant for modeling cosmology almost universally feature axion fields.  In the absence of certain fluxes that induce axion masses, such fields enjoy all-orders shift symmetries, and acquire mass only nonperturbatively. For this reason, many classes of constructions contain ultralight axions.  The number of axion fields is determined by the topology of the internal space, and the rich topologies found in contemporary constructions often give rise to large numbers of axions: hundreds, in the case of Calabi-Yau threefolds.  These findings led to the notion of the \emph{string axiverse} \cite{Arvanitaki:2009fg}, which is the idea that the low-energy effective theories resulting from quantum gravity will
contain many axions distributed over a wide range of mass scales.

Experimental and observational searches for axions are proliferating.  Even if no evidence for axions is found, these experiments will in time place limits over a wide parameter range.  With sufficient effort on the theory side, near-future advances in axion experiment may serve to
discriminate empirically among qualitatively different classes of string vacua, and to exclude a range of constructions.
In this connection, we stress an important difference between
the scenarios reviewed in \S\ref{sec:power-law} compared to those in \S\ref{sec:NP}.
In the vacua of \S\ref{sec:power-law}, perturbative effects with characteristically polynomial dependence on the canonically-normalized fields control the vacuum structure, and four-dimensional supersymmetry plays no role.  The generic choices of fluxes and branes invoked in this context typically induce a potential for axions, via monodromy.  As a consequence, even though \emph{axions} are abundant in this setting, ultralight axions need not be.

In contrast, in the vacua of \S\ref{sec:NP}, which involve Calabi-Yau compactifications with moduli stabilized by perturbative and nonperturbative effects, the imprint of supersymmetry is strong, and the leading potential energy terms for some classes of axions result from instanton contributions to the superpotential.
Moreover, characteristic trends in the sizes of cycles in Calabi-Yau manifolds with many moduli cause the typical instanton actions to be large whenever the $\alpha'$ expansion is under control \cite{Demirtas:2018akl}, and in this sense ultralight axions --- with masses $m \ll 10^{-33}\,\text{eV}$ --- are generic in the Calabi-Yau compactifications of \S\ref{sec:NP}, though not necessarily in string theory as a whole.

To capitalize on the connection to axion experiment,
a first step is to construct ensembles of many-axion effective theories resulting from string compactifications --- such theories are termed axiverses --- in enough detail so that the theory unknowns do not wash out potential signatures.
In the case of nonperturbatively-stabilized Calabi-Yau compactifications, considerable effort has gone into constructing axiverses and exploring their phenomenology.
Early works include \cite{Acharya:2010zx,Cicoli:2012sz},
while advances in computation (e.g.,~\cite{CYTools}) have allowed the enumeration of large ensembles of axiverses in Calabi-Yau compactifications \cite{Demirtas:2018akl} (more recent statistical analyses include \cite{Broeckel:2021dpz}).
A general framework for analyzing the resulting axion landscapes, and a description of their phenomenology, appears in \cite{Bachlechner:2018gew}.

Evidence for or against an axiverse may come through gravitational wave signals \cite{Kitajima:2018zco}, a multitude of cosmic strings \cite{Agrawal:2019lkr,March-Russell:2021zfq}, or the
evaporation of primordial black holes \cite{Calza:2021czr}.
Future constraints on axion dark matter
should probe some of the parameter space of string-theoretic models, and theories with ultralight axions, including Calabi-Yau compactifications, may provide candidates \cite{Cicoli:2021gss} for fuzzy dark matter \cite{Hui:2016ltb}.
Nonlinear dynamics of multi-axion dark matter \cite{Cyncynates:2021yjw} could allow detection of a small sub-component.

Direct couplings of axions to the visible sector are tested by a range of astrophysical observations and terrestrial experiments.  In some parameter ranges, constraints from gamma ray and X-ray astronomy, such as \cite{Marsh:2017yvc}, are stronger than the limits obtained from laboratory tests of couplings to the electron \cite{Ferreira:2022egk}.
Moreover, helioscope limits already constrain a subset of models with many ultralight axions \cite{Halverson:2019cmy}.  An imprint of axions in cosmic birefringence is another intriguing possiblity \cite{Komatsu:2022nvu}.

The original motivation for introducing axions in quantum field theory was the strong CP problem.
The Peccei-Quinn mechanism  \cite{Peccei:1977hh,Peccei:1977ur} is sensitive to Planck-scale physics \cite{Holman:1992us,PhysRevD.46.539,Kamionkowski:1992mf}, so a complete solution requires information from quantum gravity.
The fact that inflation depends on Planck-scale physics has been a key motivation for understanding inflation in string theory (see e.g.~\cite{Baumann:2014nda}), and for expecting that probes of inflation could teach us about string theory.  In just the same way, the strong CP problem calls out for treatment in string theory.
In the Calabi-Yau orientifold compactifications of \cite{Demirtas:2018akl}, the leading quantum-gravity contributions to the neutron electric dipole moment are due to Euclidean D3-branes, and can be computed explicitly \cite{Demirtas:2021gsq}.
When the number of axions is $\gtrsim 20$, the rich topology of the internal space leads to large cycle sizes, and the Euclidean D3-brane corrections to the
neutron electric dipole moment become negligibly small, restoring the Peccei-Quinn solution of the strong CP problem \cite{Demirtas:2021gsq}. This axiverse incarnation of the Peccei-Quinn mechanism can be probed through limits on axion dark matter and on axion emission in stars.

A particularly promising test of the axiverse picture comes from black hole superradiance.
Near a Kerr black hole, there is an instability to production of axions in a certain mass range, which can gradually spin down the black hole.  This process can occur in the vacuum, and because it does not require axions to be present as a cosmological population, the resulting limits are insensitive to assumptions about cosmic history.
Equipped with the distribution of axion masses in axiverses arising in Calabi-Yau compactifications, one can use the measured spins of astrophysical black holes to place limits on these solutions \cite{Mehta:2021pwf}, excluding some regions of moduli space and some ranges of topological parameters.

A comprehensive review of axion phenomenology is beyond the scope of this white paper, and we have highlighted only a few of the more recent connections to string theory. Further context can be found in the review \cite{Marsh:2015xka} and the white paper \cite{axionwhitepaper}.

\section{Connections to other areas}\label{sec:connections}

In this final section we describe several connections between the theory frontier in cosmology and other areas of inquiry.

\subsection{Mathematics}

There are many links between mathematics and inflationary physics.  In \S\ref{sec:classical-initial} we highlighted the mathematical proofs of the robustness of inflation.  It is also notable that only in the de Sitter case has a full proof of stability of the Kerr black hole been achieved \cite{10.4310/ACTA.2018.v220.n1.a1}. Integrability-based methods in mathematical physics such as the exact solvability of certain deformations of quantum field theories extend readily to the case with positive cosmological constant, as detailed above in \S\ref{sec:holography}.

Compactification geometry and topology, as described in \S\ref{sec:string-inflation}, is another major area of intersection.   Methods include group theory, differential geometry, algebraic geometry, and topology, depending on the setting.
Hyperbolic manifolds and more general negatively curved spaces are well-studied mathematically.
As stressed earlier, they are also well-motivated physically given their positive contribution to the potential, and their rigidity \cite{Besse:1987pua}; the latter combined with warp factor dynamics \cite{Douglas:2009zn} contributes a strong positive Hessian for metric deformations.  In particular, such manifolds do not have moduli in the mathematical sense, analogously to the absence of exactly flat directions in most quantum field theories. Powerful methods exist to analyze them \cite{Ratcliffe:2006bfa} with the help of simple examples \cite{italiano2020hyperbolic}.  Another timely theme concerns systolic geometry
\cite{AgolInbreeding, Thomsonethesis}, relating the length of the minimal closed geodesic to the overall size of the manifold.  This UV-IR effect has interesting physical implications \cite{DeLuca:2021pej}.  The infinite sequences of hyperbolic spaces contrasts with the conjecturally finite list of Calabi-Yau threefolds, and enables parametric tunings of control parameters \cite{Saltman:2004jh, DeLuca:2021pej}.  These spaces participate in string dualities and connect to other parts of the landscape \cite{McGreevy:2006hk, Green:2007tr, Adams:2005rb}\ in ways tied to their rich topology (e.g. the exponential growth of their fundamental group). Their generality, positivity, and rigidity fits with current empirical constraints on particle physics and cosmology, motivating additional study.
In the case of Calabi-Yau   compactifications, a recent highlight is the analytic derivation of K3 metrics \cite{Kachru:2020tat}.  At the same time, constructions of cosmological vacua based on Calabi-Yau compactifications have required advances in computational algebraic geometry \cite{CYTools}.  The resulting tools have revealed new structures in the geometry and topology of the underlying manifolds.

\subsection{Machine learning}\label{sec:ML}

The enormity of the string landscape makes exhaustive searches or random sampling impractical.
Even restricting to presently-known classes of Calabi-Yau threefolds, the number of distinct vacuum geometries may be as large as $10^{428}$ \cite{Demirtas:2020dbm}, while counting choices of fluxes in Calabi-Yau compactifications gives as many as $10^{500}$ \cite{Ashok:2003gk,Denef:2004ze}  to $10^{272,000}$~\cite{Taylor:2015xtz} flux vacua.
As noted above, other types of compactifications come in infinite sequences of topologies and dimensionalities at bounded Kaluza-Klein scale.
The vastness of the landscape is exacerbated by computational complexity.  Finding specific string vacua appears to be NP-hard in many cases \cite{Denef:2006ad,Denef:2017cxt,Halverson:2018cio}, though
this measure of complexity is based on brute force search, which can be vastly outperformed by optimization algorithms \cite{Bao:2017thx}.\footnote{Limitations from computational complexity can also be evaded in constructions such as \cite{Demirtas:2019sip,Demirtas:2021nlu}, where the computational cost of enumerating vacua indeed appears to be exponential in a dimension-counting parameter $N$ (corresponding in \cite{Demirtas:2019sip,Demirtas:2021nlu} to the Hodge number $h^{2,1}$ of a Calabi-Yau threefold), but viable solutions already appear at $N=2$.}

Only a minute subset of the configurations mentioned above are expected to be phenomenologically viable.
In search of realistic vacua, additional criteria are necessarily imposed --- either to set phenomenological properties, such as the observed vacuum energy and particle spectrum, or to ensure theoretical control, for example with small couplings and large volumes.
These additional criteria put further structure on the landscape, with the optimization target being solutions of some constrained systems of equations. The highly ordered structure of the landscape together with the discreteness of the UV inputs (e.g, fluxes and topology of the compactification) leads to a rich topological structure of voids and voxels in the landscape \cite{Cole:2018emh}.

Given the scope and nature of the problem, it is natural to leverage advances in machine learning to design efficient optimization methods to search for realistic vacua that may at the same time reveal hidden structure in the landscape. The question of interest is not only the construction of realistic vacua but also the statistical distribution surrounding such vacua, which has implications for various dynamics-based proposals for the measure problem \cite{Denef:2017cxt,Bao:2017thx,Khoury:2019yoo}. In recent years, stochastic optimization with Genetic Algorithms (GAs) \cite{Blaback:2013ht,Blaback:2013fca,Abel:2014xta,Cole:2019enn,AbdusSalam:2020ywo,Bena:2021wyr, Loges:2021hvn} and Reinforcement Learning (RL) \cite{Halverson:2019tkf,Larfors:2020ugo,Krippendorf:2021uxu,Constantin:2021for} have been utilized to search for viable string vacua, outperforming searches based on Metropolis-Hastings.
A first comparison between GAs and RL in a string theory context  \cite{Cole:2021nnt} showed that both broad approaches are effective in searching for optimal vacua and discovering structures of the landscape, but in complementary ways.
{ In general, one may work from the formalism \eqref{VeffEframe-string} \cite{Douglas:2009zn} to explore the effective potential, or descend in its functional derivatives related to slow roll parameters, starting from a particular topology and fiducial geometry along with calculable stress-energy sources such as Casimir energy, brane tensions and the like.}

More broadly, a main goal of cosmology is to reconstruct the initial inhomogeneities of the universe from current observations.
Primordial non-Gaussianity provides valuable probes of the interactions \cite{Alishahiha:2004eh,Chen:2006nt,Moss:2007cv,LopezNacir:2011kk} and particle content \cite{Chen:2009zp,Baumann:2011nk,Assassi:2012zq,Noumi:2012vr,Arkani-Hamed:2015bza} during inflation, which can help break the degeneracy of models.
Unfortunately, the physical evolution from the primordial matter distribution to the observations today is extremely complicated and non-linear. Different theories of the early universe predict tiny non-Gaussian features in the primordial matter distribution, which underwent 13.8 billion years of astrophysical evolution. Upcoming experiments will result in massive new datasets, but it is not known whether they can be modeled precisely enough to probe these non-Gaussianities. Machine learning (ML), with its ability to model highly non-linear phenomena, could be the crucial missing ingredient to make this possible. Topological data analysis (TDA) is an ML and statistical method for summarizing the shape of data. TDA extracts non-local information of position-space maps, which can distinguish primordial non-Gaussianity from the non-linearities of gravity  \cite{Cole:2017kve, Biagetti:2020skr}. More generally, it was argued in \cite{Baumann:2021ykm} that a map-level analysis may significantly improve the constraining power over previous forecasts.
Thus, ML approaches such as simulation-based inference involving the forward modeling of large-scale structure maps \cite{Schmittfull:2017uhh,Taylor:2019mgj,Dai:2020ekz,Modi:2020dyb,Modi:2021acq,Makinen:2021nly,Hassan:2021ymv,Villaescusa-Navarro:2021pkb,Villaescusa-Navarro:2021cni, Cole:2021gwr} may have the potential to dramatically impact the search for primordial non-Gaussianity.

ML also finds applications in computing Calabi-Yau metrics numerically  \cite{Anderson:2020hux,Douglas:2020hpv, Jejjala:2020wcc}
and more generally in solving partial differential equations \cite{DeLuca:2022brp} arising in string compactifications.
While many quantities of interest in Calabi-Yau compactifications can be computed without explicit knowledge of the internal metrics, thanks to Yau's theorem, this is not the case for the general compactifications outlined in \S\ref{sec:power-law}. Moreover, even for Calabi-Yau compactifications, non-holomorphic quantities --- including the inflation potential and the kinetic terms of fields ---
are not amenable to algebraic geometry methods.  For example, computation of inflation potentials in D-brane inflation so far utilizes local Calabi-Yau metrics \cite{Baumann:2007ah} and parametrizes models in terms of boundary conditions compatible with these metrics' approximate isometries \cite{Baumann:2008kq,Baumann:2010sx,Gandhi:2011id}.
Further advances in ML techniques for computing compact Calabi-Yau metrics may allow us to determine the boundary conditions and quantify the breaking of local isometries, as in \cite{Aharony:2005ez}. Such advances are also needed for addressing metric-dependent questions in D-brane inflation, such as the process of reheating \cite{Kofman:2005yz,Chialva:2005zy,Chen:2006ni}. In the case of the negatively curved compactifications discussed in \S\ref{sec:power-law}, the fiducial (hyperbolic) metric is known, in contrast to the case of Calabi-Yau manifolds. There it is interesting to move on to numerically analyze the internal PDEs for the warp and conformal factors using a neural network
as a function ansatz, as was done as a proof of principle in a hyperbolic polygon in \cite{DeLuca:2021pej} (though analytic and simpler numeric methods sufficed so far).     Aside from classical PDEs, interesting connections between quantum fields and neural networks are under active development (see e.g. \cite{deHaan:2021erb,Halverson:2021aot}).

Finally, we may turn this connection around and apply ideas from cosmology to design algorithms for machine learning and other numerical methods.  This approach is related to a very active general program of research applying physics methods to improve and better understand the process, see e.g.~\cite{roberts2021principles}. A new class of optimizers under development \cite{DeLuca:2022brp}\ was inspired by the field-dependent speed limit in \cite{Alishahiha:2004eh}; identifying the squared speed limit with the loss function and working in the non-gravitational, energy-conserving limit yields a frictionless dynamics that nonetheless slows the evolution near vanishing loss.  The energy conservation in this and many similar examples yields some favorable properties and calculable predictions for the performance, with large phase space measure near the desired regime of small loss.  A future direction is to flesh out this and other cosmology-based ideas in large-scale numerical problems.

\section*{Acknowledgements}

The work of R.F.~was supported in part by the Department of Energy under Grant No.~DE-SC0009919, the Simons Foundation/SFARI 560536, and by NASA under Grant No.~80NSSC18K1487 and 80NSSC18K0561. The work of L.M.~was supported in part by
National Science Foundation grant PHY-1719877.
The work of G.S.~was supported in part by Department of Energy grant DE-SC0017647.
The work of E.S.~was supported in part by a Simons Investigator award and National Science Foundation grant PHY-1720397.

\bibliographystyle{JHEPsnowmass}
\bibliography{refs}
\end{document}